\documentclass[nofootinbib,amsmath,amssymb,groupedaddress,aps,pra]{revtex4-2}

\usepackage{graphicx}

\begin{document}


\title{Damped harmonic oscillator revisited: a new approach to energy decay in the case of Coulomb, Stokes, and Newton damping}

\def\correspondingauthor{\footnote{Corresponding author: rpezer@simet.unizg.hr}}
\author{Robert Pezer\correspondingauthor{}}
\affiliation{Department of Physical Metallurgy, Faculty of Metallurgy, University of Zagreb, Aleja narodnih heroja 3, 44000 Sisak, Croatia}
\email{rpezer@simet.unizg.hr}

\author{Karlo Lelas}
\affiliation{Faculty of Textile Technology, University of Zagreb, Prilaz baruna Filipovića 28a, 10000 Zagreb, Croatia}
\email{klelas@ttf.unizg.hr}


\date{\today}

\begin{abstract}
Approximate formulas are derived to describe energy loss in a harmonic oscillator that experiences three distinct damping mechanisms: constant-magnitude (Coulomb), velocity-proportional (Stokes), and velocity-squared (Newton), using fundamental mathematical methods and physical insight. Our methodology leverages an understanding of the free harmonic oscillator and the inherent link between energy dissipation rates and the power exerted by damping forces. We establish a direct analytical framework for assessing the energy of a damped harmonic oscillator, obviating the need for amplitude-based equations. The simplicity of our findings is accompanied by their remarkable accuracy when validated against exact or computational simulations. In addition to an excellent approximate description of the energy decay, we also show how to derive an exact solution in the case of Stokes damping without relying on the standard procedure for solving second-order differential equations. The theoretical underpinnings and mathematical strategies employed are well-suited for undergraduate-level or advanced high school physics instruction.
\end{abstract}


\maketitle

\section{Introduction \label{sec:intro}}

Teaching the damped harmonic oscillator (DHO) in an undergraduate physics course requires balancing theoretical rigor and physical intuition, as it is a foundational concept in mechanics and oscillatory systems. The harmonic oscillator (HO), without and with damping, is taught to undergraduate physics \cite{HallidayResnick10,Young2020university} and engineering \cite{Hartog1974MV} students because it provides a simple yet powerful model for understanding oscillatory motion, energy principles, and dynamic systems. The undamped case builds foundational skills in physics and mathematics, while the damped case introduces realistic energy dissipation, nonlinear dynamics, and engineering applications. Together, they equip students with analytical, computational, and design skills, preparing them for advanced studies and practical challenges in mechanical, civil, electrical, and aerospace engineering. The interdisciplinary relevance of the topic, along with its real-world applications, makes it essential for fostering both theoretical understanding and practical problem-solving.
The cornerstone feature of this system is the conversion of mechanical energy between potential and kinetic forms. The real beauty of the DHO lies in its unique combination of relative simplicity and richness in physical features, particularly in terms of energy.
In recent papers, some interesting new insights \cite{Lelas2023, LelasPezerEJP2024Mod, LelasPezer2025Add} arise, and here we continue along these lines.

In this paper, we analyze three classic forms of dissipative forces separately in the context of a harmonic oscillator:
\begin{enumerate}
	\item Coulomb (Constant) Damping, force law: $F_d = -m c_0\, sgn(v)$, independent of speed, nonlinear;
	\item Stokes (Linear) Damping, force law: $F_d = -m c_1 v$;
	\item Newton (Quadratic) Damping, force law: $F_d = -m c_2\, sgn(v)\, v^2$, nonlinear;
\end{enumerate}
where we used the signum function with properties $sgn(v)=1$ for $v>0$, $sgn(v)=-1$ for $v<0$, and $sgn(v)=0$ for $v=0$.
That said, we take the opportunity to study many essential concepts in a simple, tractable context, including energy dissipation, linearity versus nonlinearity, the period of motion, analytical reasoning about decay forms, and the physics of dissipation from an energy perspective.

The following is a summary of the primary strategies for effectively teaching this topic at the undergraduate engineering level, with a focus on clarity, engagement, and conceptual understanding.

In this paper, we present a quantitative analysis of the dissipation of the total DHO mechanical energy (here expressed with usual dimensional quantities)
\begin{equation} \label{eq:Emeh} 
	E_{m}(t) = \frac{1}{2} m \, \dot{x}_{d}^2(t) \, + \, \frac{1}{2} k \, x_{d}^2(t) = E_k(t) + E_p(t)
\end{equation}
using only the notion of elementary function derivative and integration. A cornerstone feature of this analysis is the time dependence of \eqref{eq:Emeh}, including the redistribution of energy between the kinetic and potential shares.

\section{Formalism \label{sec:formalism}}
The HO with damping up to the square of velocity leads to a dynamic differential equation:
\begin{equation} \label{eq:Newton2nd4x} 
	-m (c_0 sgn( \dot{x}_{d} ) + c_1 \dot{x}_{d} + c_2\, sgn( \dot{x}_{d}) \dot{x}_{d}^2 ) - kx_{d} = m\ddot{x}_{d} 
\end{equation}
Dividing both sides by the mass m and rearranging, we get:
\begin{equation*}
	\ddot{x}_{d} + c_0 sgn( \dot{x}_{d} ) + c_1 \dot{x}_{d} 
	+ c_2\, sgn( \dot{x}_{d}) \dot{x}^2 + \omega_0^2 \, x_{d} = 0 
\end{equation*}
where we have used the usual temporal parameters:
$$ \omega_0^2 = k/m \qquad T_0 = 2\pi/\omega_0 $$
Appendix \ref{app:DHO_lin_sol} offers a comprehensive solution for Stokes damping ($c_1 \ne 0 $ and $ c_0=c_2=0$) by relying solely on the basic differentiation rules applied to elementary functions. This approach completely bypasses the complexities associated with solving second-order differential equations.

Let us now rewrite Eq. \eqref{eq:Newton2nd4x} with friction force
\[ F_{drag} = -m (c_0 sgn( \dot{x}_{d} ) + c_1 \dot{x}_{d} + c_2\, sgn( \dot{x}_{d}) \dot{x}_{d}^2 )   \]
in terms of energy quantities.
We are starting from a well-known expression for the mechanical energy dissipation rate:
\begin{equation}
	\frac{d E_m}{dt} =  \dot{x}_{d} \,F_{drag} = 
	-m \left[c_0\, sgn( \dot{x}_{d} )\, \dot{x}_{d} + c_1\, \dot{x}_{d}^2 + c_3\, \left(sgn( \dot{x}_{d})\, \dot{x}_{d}\right)^3 \right]
	\label{eq:Em_Fdrag}
\end{equation}
We note that Eq. \eqref{eq:Em_Fdrag} could be derived as the rate of mechanical energy \eqref{eq:Emeh} decay by multiplying both sides of \eqref{eq:Newton2nd4x} by $\dot{x}_{d}$ and gathering terms appropriately.

It is convenient to notice the following expression relating velocity to kinetic energy
$E_k$ (always a positive quantity):
\begin{equation}
	sgn( \dot{x}_{d} )\, \dot{x}_{d} = \sqrt{\frac{2 E_k}{m}}
	\label{eq:K_v}
\end{equation}
enabling us to formulate the energy-based equation derived from Eq. \eqref{eq:Newton2nd4x}.
After substitution \eqref{eq:K_v} in \eqref{eq:Em_Fdrag} and straightforward transformation:
\begin{equation*}
	\frac{d E_{m}}{dt} =  - c_0 m \left(\frac{2 E_k}{m} \right)^\frac{1}{2}
	- c_1 m \frac{2 E_k}{m} - c_2 m \left(\frac{2 E_k}{m} \right)^\frac{3}{2}
\end{equation*}
\begin{equation} \label{eq:Emech_rate}
	\frac{d E_{m}}{dt} =  - c_0 \left(2 m E_k \right)^\frac{1}{2}
	- 2 c_1 E_k 
	- 2 c_2 E_k \left(\frac{2 E_k}{m} \right)^\frac{1}{2}
\end{equation}
we arrive at the first-order differential equation for energy.
Therefore, the dissipation rate is expressed solely in terms of energy quantities. The trouble is, we have different functions on the left and right side of Eq. \eqref{eq:Emech_rate}, i.e. $E_m(t)$ and $E_k(t)$. We solved the Stokes problem exactly in appendix \ref{app:DHO_lin_sol}, where the exponential mechanical energy decay envelope appears. For two other cases considered here, we seek good approximations that capture the oscillating part to some extent, and not only the decaying part of the energy as was done in \cite{WangEJP2002,LelasPezer2025AmpDec}.
The Coulomb case is exactly solvable \cite{Marchewka2004Coulomb, Kamela2007Coulomb, AnastasiosAdamopoulosa2022Coulomb}, giving rise to second-degree polynomial decay, while Newton damping approximately follows an inverse second-degree polynomial decay pattern \cite{WangEJP2002}.

Without compromising the physical content, we consider the initial condition $E_k(0) = 0$ (i.e., $x_{d}(0) = x_0>0$ and $\dot{x}_{d}(0) = 0$).
Let us, in addition, introduce dimensionless time and energy quantities:
\begin{equation} \label{eq:E_reduced}
	\tau = \omega_0 t \quad
	\mathcal{E}_l (\tau) =  \frac{E_l}{\frac{1}{2} m (\omega_0 x_0)^2} \quad l= m,k,p
\end{equation}
Introducing these quantities, we can rewrite Eq. \eqref{eq:Emech_rate} in the fully dimensionless form
\begin{equation} \label{eq:e_tot_rate}
	\frac{d \mathcal{E}_{m}}{d\tau} =  
	- 2\gamma_0 \left(\mathcal{E}_k \right)^\frac{1}{2}
	- 2\gamma_1 \mathcal{E}_k 
	- 2\gamma_2 \left(\mathcal{E}_k \right)^\frac{3}{2}
\end{equation}
where
\[
\gamma_0 = \frac{c_0}{\omega_0^2 x_0} \quad
\gamma_1 = \frac{c_1}{\omega_0} \quad
\gamma_2 = c_2 x_0
\]
As a brief observation, we note that in the Newton damping case, $\omega_0$ is a pure timescale (the dynamics is unaffected by this parameter), in contrast to the Coulomb and Stokes cases.

How does Eq. \eqref{eq:e_tot_rate} enable us to characterize the energy dissipation phenomena occurring under conditions of Coulomb, Stokes, or Newton drag? First, we note the problem with Eq. \eqref{eq:e_tot_rate}. The equation mixes $\mathcal{E}_{m}(\tau)$ and $\mathcal{E}_k(\tau)$, so, we need another equation specific to the physics involved connecting $\mathcal{E}_{m}(\tau)$ and $\mathcal{E}_k(\tau)$. Apart from that, Coulomb and Newton drag cases are nonlinear, while the Coulomb damping case is analytically solvable \cite{Lapidus1970AJP,Marchewka2004Coulomb, Kamela2007Coulomb, AnastasiosAdamopoulosa2022Coulomb}; there is no known exact closed-form solution for Newton damping \cite{Smith2012Newton, AnastasiosAdamopoulosa2022Coulomb}. Since we are interested in the overall and per-cycle mechanical energy loss here, we can start from the weak-damping limit and conjecture a relation between the total and kinetic energies.
The approach we will use, along with all the derivations, can be easily adapted to any other choice of initial conditions. For this choice of initial conditions, the undamped harmonic oscillator (i.e., the case with $\gamma_0=\gamma_1=\gamma_2=0$) has kinetic energy $\mathcal{E}_k(\tau)=\sin^2(\tau)$ and total mechanical energy $\mathcal{E}_{m}(\tau)=1$. Thus, in the undamped case, the ratio of kinetic to total mechanical energy is $\mathcal{E}_k(\tau)/\mathcal{E}_{m}(\tau)=\sin^2(\tau)$. It is reasonable to assume that for weak damping (i.e., for $\gamma_0\ll1$, $\gamma_1\ll1$, and $\gamma_2\ll1$), a good approximation of this ratio will be given by:
\begin{equation} \label{eq:EkEm_ratio}
	\frac{\mathcal{E}_{k}(\tau)}{\mathcal{E}_{m}(\tau)} \simeq \sin^2 (\tau)
\end{equation}
Expression \eqref{eq:EkEm_ratio} finds support in the analysis of the exact energy under Stokes damping conditions (cf. Eq. \eqref{eq:EkEm_ratio_ApproxApp} in appendix A), which concurrently provides a compelling basis for its application in estimating energy decay for the other two damping modalities.

\begin{figure}[t]
	\includegraphics[height=7cm, angle=-0]{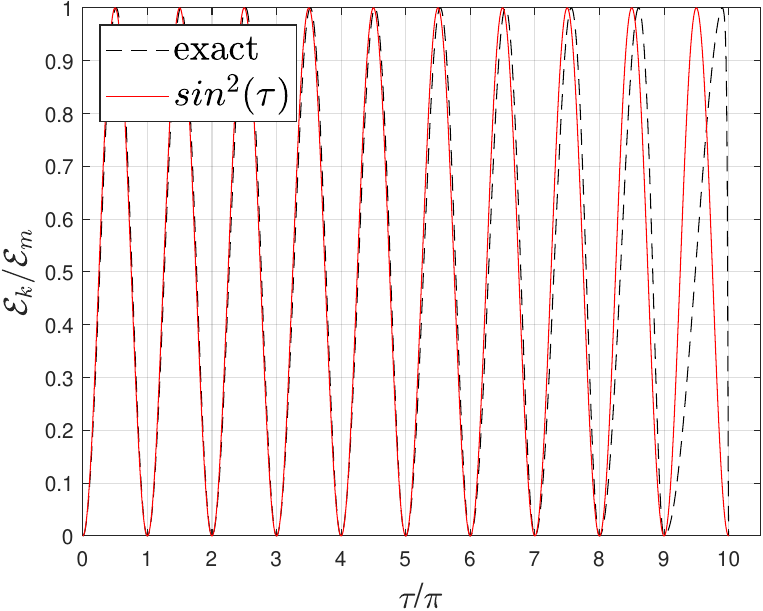}
	\includegraphics[height=7cm, angle=-0]{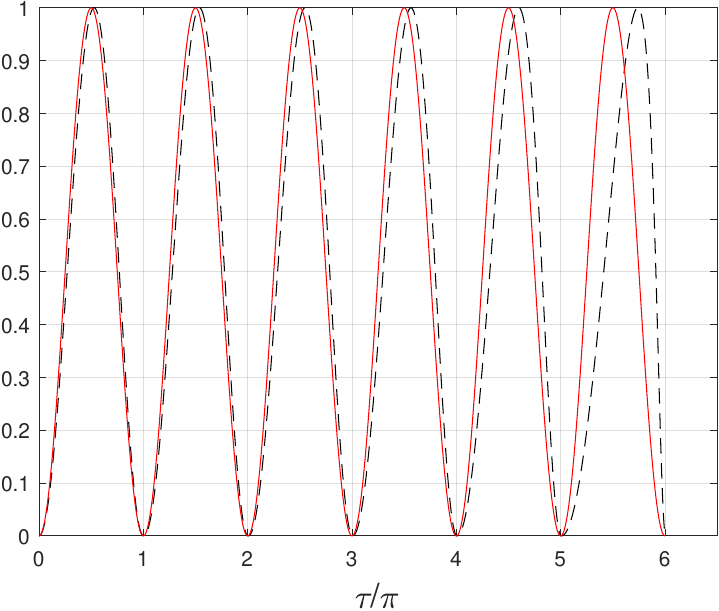}
	\caption{Coulomb damping, kinetic to total mechanical energy ratio, comparison of the exact result (dashed black line) and approximation (full red line) representing Eq. \eqref{eq:EkEm_ratio} for damping strengths $\gamma_0 = 0.050$ (left pane) and $\gamma_0 = 0.080$ (right pane). \label{fig:Ek_r_E0_Coulomb}}
\end{figure}
\subsection{Coulomb damping \label{ssec:D_Coulomb}}
Let us estimate mechanical energy loss under constant friction during a first semi-interval $[0,\pi]$, 
assuming the elastic force is strong enough to overcome static friction and set the system in motion.
The initial conditions are $\mathcal{E}_{m}(0) = \mathcal{E}_{p}(0)=1$ and $\mathcal{E}_{k}(0)=0$. Still, we stick with this choice to keep the formalism as simple as possible.
Using Eq. \eqref{eq:EkEm_ratio} in the energy dissipation rate \eqref{eq:Emech_rate} we get:
\begin{equation} \label{eq:Ep_r_Em}
\frac{d \mathcal{E}_{m}(\tau)}{d\tau} =  
- 2\gamma_0 \left(\mathcal{E}_k(\tau) \right)^\frac{1}{2} \simeq - 2\gamma_0 \sqrt{\mathcal{E}_{m}(\tau) \sin^2(\tau)} 
  = - 2\gamma_0 \sqrt{\mathcal{E}_{m}(\tau) }\, | \sin(\tau) |
\end{equation}
We further assume that the initial elongation is large enough to initiate motion (the elastic force exceeds static friction).
The accuracy of this approximation is demonstrated in Fig. \ref{fig:Ek_r_E0_Coulomb}. In this manner, in addition to using approximations, we incorporated physics specific to the system into our equations. 
As we see in Fig. \ref{fig:Ek_r_E0_Coulomb}, the maxima of the exact ratio $\mathcal{E}_{k}(\tau)/\mathcal{E}_{k}(\tau)$ lag the maxima of the approximate ratio, i.e. of $\sin^2(\tau)$, and the effect increases as the oscillation amplitude decays, i.e. towards the end of the motion.
\begin{figure}[b]
	\includegraphics[width=0.45\textwidth, angle=-0]{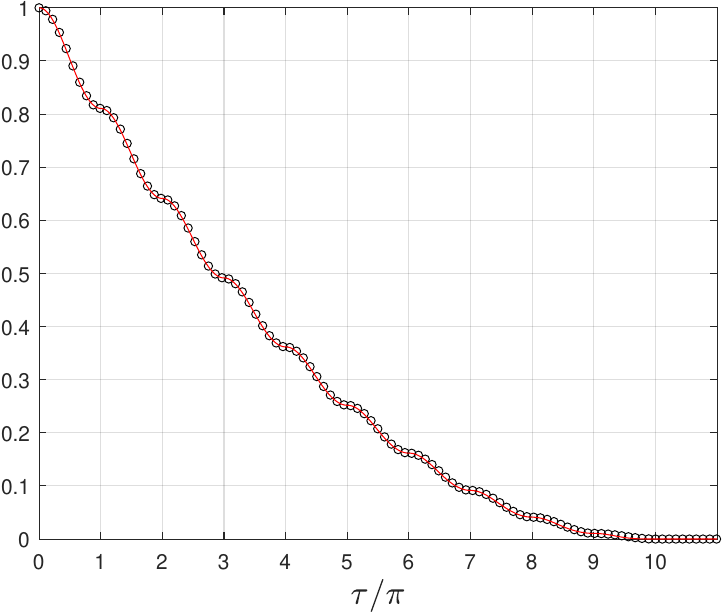}
	\includegraphics[width=0.45\textwidth, angle=-0]{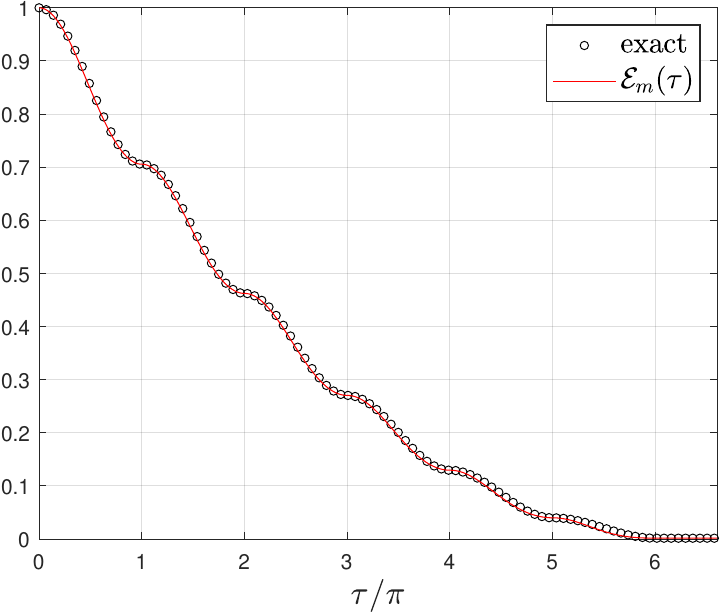}
	\caption{Coulomb damping, mechanical energy decay, exact (circles) and approximation (full red line) shown here for damping strengths $\gamma_0 = 0.050$ (left) and $\gamma_0 = 0.080$ (right). \label{fig:Etot_Coulomb_app}}
\end{figure}

Solving the differential equation \eqref{eq:Ep_r_Em} is straightforward. We have two solutions depending on the sign of $\sin(\tau)$:
\[
\sqrt{\mathcal{E}_{m}(\tau)} = \gamma_0 \cos(\tau) + C_e \quad 0 \le \sin(\tau) \Leftrightarrow \tau \in [2n\pi, (2n+1)\pi]
\]
\[
\sqrt{\mathcal{E}_{m}(\tau)} = -\gamma_0 \cos(\tau) + C_o \quad 0 > \sin(\tau) \Leftrightarrow \tau \in [(2n+1)\pi,2(n+1)\pi]
\]
\[n=0,1,2,3, \ldots, n_{max}\]
where $n_{max}$ is yet unknown maximal number consistent with physical constraints, and integration constants $C_e$ and $C_o$ are determined by the initial condition $\mathcal{E}_{m}(0) = 1$ and the continuity of $\mathcal{E}_{m}(\tau)$ at the instants $\tau=(n+1)\pi$, where we transition from one solution to another. It is straightforward to show that the final approximate mechanical energy is:
\begin{equation} \label{eq:C_Em_app_g}
	\mathcal{E}_{m} (\tau) = \left[1 - \gamma_0 \left(2n+1 - (-)^n \cos(\tau) \right)\right]^2 
	\qquad \tau \in \pi [n,n+1], n=0,1,2,3,\ldots, n_{max}
\end{equation}
In Fig. \ref{fig:Etot_Coulomb_app}, we show examples of Eq. \eqref{eq:C_Em_app_g} for two characteristic damping strengths. We see that the approximation captures all the qualitative features of the exact dynamics, and its quantitative accuracy is remarkable as well.

\begin{figure}[t]
	\includegraphics[width=0.45\textwidth, angle=-0]{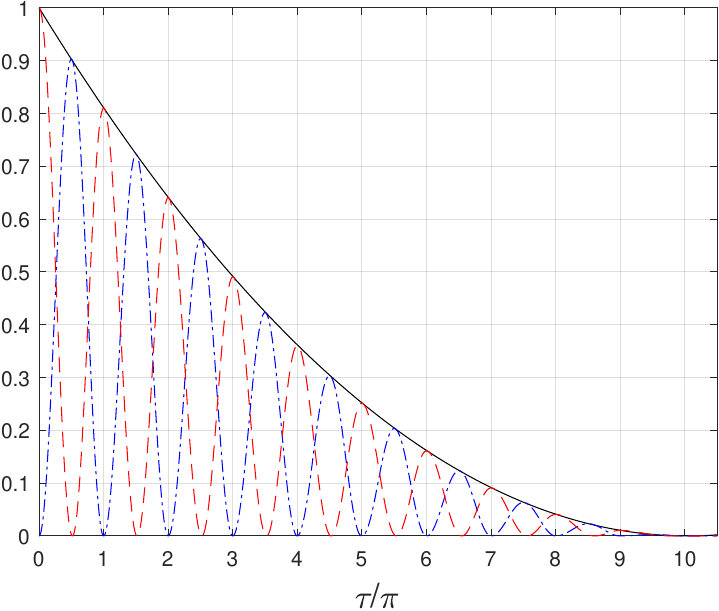}
	\includegraphics[width=0.45\textwidth, angle=-0]{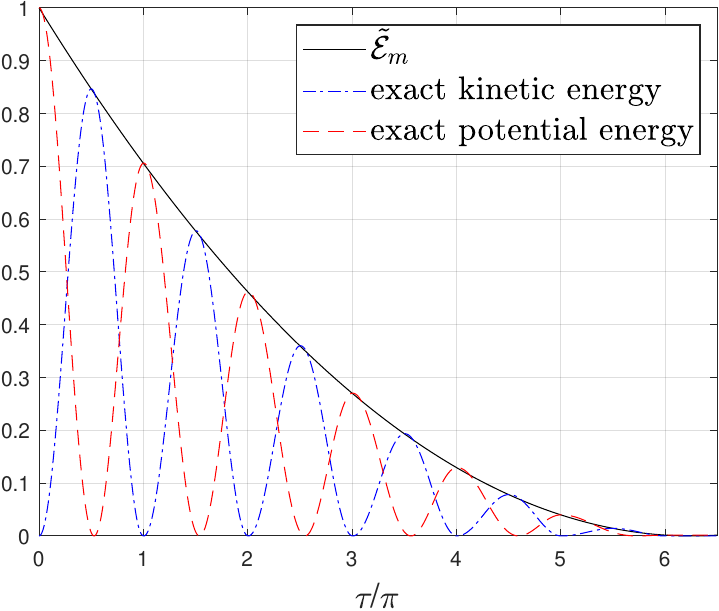}
	\caption{Coulomb damping, approximate envelope $\mathcal{\tilde{E}}_{m}$ (solid black line, Eq. \eqref{eq:approxEm_Coulomb}) of both, exact kinetic (dot dashed line) and potential (dashed line) energy. Comparison is given for two damping strengths $\gamma_0 = 0.050$ (left) and $\gamma_0 = 0.080$ (right). \label{fig:Etot_Coulomb}}
\end{figure}
The approximate energy $\mathcal{E}_{m} (\tau)$ consists entirely of the potential energy at instants $\tau_n=n\pi$, for $n\in\lbrace0,1,2,\dots,n_{max}\rbrace$, since the kinetic energy is equal to zero at instants $\tau_n$ within our approximation, due to $\sin(\tau_n)=0$. Using $\tau=\tau_n$ and $\cos(\tau_n)=(-1)^n$ in \eqref{eq:C_Em_app_g}, we easily get
\begin{equation} \label{eq:envelope0_E_Coulomb}
	\mathcal{E}_{m} (\tau_n) = \left(1 - 2n\,\gamma_0\right)^2=\left(1 - \frac{2}{\pi}\gamma_0 \tau_n\right)^2\,,
\end{equation}
where we formally wrote $n=\tau_n/\pi$ after the second equality to explicitly show how the local maxima of the potential energy decay over time. 
Thus, we can easily conclude that the envelope of the potential as a function of time is approximately given by 
\begin{equation} \label{eq:envelope_E_Coulomb}
	\mathcal{\tilde{E}}_{p} (\tau) = \left(1 - \frac{2}{\pi}\gamma_0\tau\right)^2\,.
\end{equation}

\begin{figure}[b] 
	\includegraphics[width=0.45\textwidth, angle=-0]{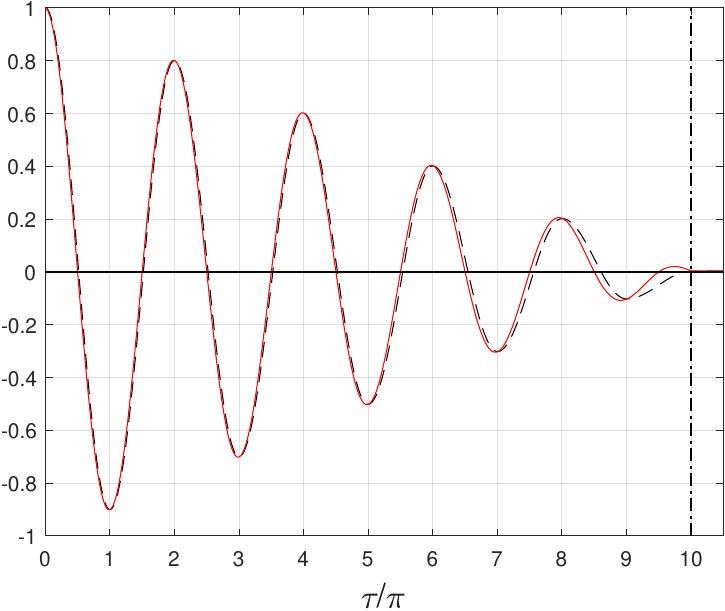}
	\includegraphics[width=0.45\textwidth, angle=-0]{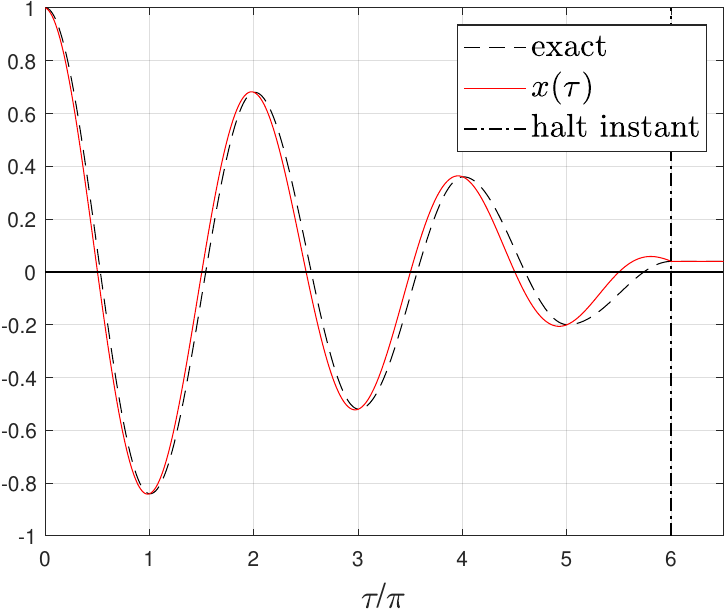}
	\caption{Coulomb damping, exact displacement (dashed black line) against approximate displacement $x(\tau)$ (solid red line) given in Eq. \eqref{eq:Coulomb_x_app}. Comparison is presented for two damping strengths: $\gamma_0 = 0.050$ (left pane) and $\gamma_0 = 0.080$ (right pane). The dot-dashed vertical line serves to indicate a pre-established instant (see Eq. \eqref{eq:coulomb_e_full}) at which motion ceases.
		\label{fig:x_Coulomb} }
\end{figure}
For an undamped harmonic oscillator, the potential energy is given by $\mathcal{E}_{p} (\tau)=\cos^2(\tau)$ (for our choice of initial conditions and units, see Eq. \eqref{eq:E_reduced}). In the case of weak Coulomb damping, the potential energy oscillates with decreasing amplitude, i.e., within the envelope approximately given by Eq. \eqref{eq:envelope_E_Coulomb}. Thus, we can approximate the potential energy as 
\begin{equation} \label{approxEP}
	\mathcal{E}_{p} (\tau) = \mathcal{\tilde{E}}_{p} (\tau)\cos^2(\tau)=\left(1 - \frac{2}{\pi}\gamma_0\tau\right)^2\cos^2(\tau)\,.
\end{equation}
The corresponding (dimensionless) approximate displacement as a function of time is given by
\begin{equation}
	x(\tau) = \sqrt{\mathcal{E}_{p} (\tau)}=\left(1 - \frac{2}{\pi}\gamma_0\tau\right)\cos(\tau)\,.
	\label{eq:Coulomb_x_app}
\end{equation}
We show a comparison of the exact and approximate displacement in Fig. \ref{fig:x_Coulomb}. In the figure, we used our knowledge of the stopping time, which will be derived and shown later in Eq. \eqref{eq:coulomb_e_full}. We note that the approximation is fairly precise even for this moderate damping (system exhibits just 3 oscillation cycles before stopping). Approximation breaks down in the final cycle because the dissipated energy becomes comparable to the residual potential energy.

Of course, once we have obtained the approximate energy $\mathcal{E}_{m} (\tau)$, i.e. \eqref{eq:C_Em_app_g}, one may ask why not determine the approximate kinetic and approximate potential energy using the ratio Eq. \eqref{eq:EkEm_ratio}. One can obtain corresponding approximations, but in this way, we would obtain piecewise-defined approximate expressions, not the closed-form functions in \eqref{approxEP} and \eqref{eq:Coulomb_x_app}.

In the same way as we derived the approximate envelope of the potential energy, but considering the instants $\tau_n=\left(n + \frac{1}{2}\right)\pi$, we obtain an approximation of the envelope of the kinetic energy 
\begin{equation*} 
	\mathcal{\tilde{E}}_{k} (\tau) = \left(1 - \frac{2}{\pi}\gamma_0 \tau\right)^2\,,
\end{equation*}
i.e. the same function as \eqref{eq:envelope_E_Coulomb}. Thus, within our approximation, the kinetic energy is 
\begin{equation} \label{eq:approxEk_Coulomb}
	\mathcal{E}_{k} (\tau) =\left(1 - \frac{2}{\pi}\gamma_0\tau\right)^2\sin^2(\tau)\,.
\end{equation}
Using Eq. \eqref{approxEP} and \eqref{eq:approxEk_Coulomb}, we can get a closed-form approximation of the total mechanical energy
\begin{equation} \label{eq:approxEm_Coulomb}
	\mathcal{E}_{m} (\tau) = \mathcal{E}_{k} (\tau)+\mathcal{E}_{p} (\tau)=\left(1 - \frac{2}{\pi}\gamma_0\tau\right)^2 \equiv \mathcal{\tilde{E}}_{m} (\tau)\,.
\end{equation}
giving us, directly, total mechanical energy envelope function as well.
The closed-form approximation of the total energy of the same form as Eq. \eqref{eq:approxEm_Coulomb} was already obtained in \cite{WangEJP2002} using the time averaging of the energy dissipation rate. We present herein an alternative, and in our assessment, a less complex derivation, which is more advantageously applied within an introductory educational settings considered here.

Let us now determine the instant at which motion halts and the residual energy left frozen in the system. We can right away start with estimate $\tau_{max} = \frac{\pi}{2 \gamma_0}$, which is clear from Eq. \eqref{eq:approxEm_Coulomb}, since for any $\tau > \frac{\pi}{2 \gamma_0}$ system's total mechanical energy $\mathcal{\tilde{E}}_{m} (\tau)$ would increase.

However, we can be even more precise about the stopping time and amplitude using straightforward simple physical reasoning.
Motion stops once the static Coulomb friction force is greater than the elastic restoring force at some potential energy maxima ($v=0$) of the amplitude. 
For motion to continue, the elastic restoring force should overcome static friction for every $n$
consistent with motion presence. Condition with dimensional physical quantities
\[ x_0\, x(\tau) = x_d(\tau/\omega_0) \]
for $n$ might be stated as:
\[
\left|  k\, x_d\left(n \frac{\pi}{\omega_0}\right) \right| > m c_{0s} \Rightarrow 
\left| \frac{x_d\left(n \frac{\pi}{\omega_0}\right)}{x_0} \right| = \left|x(n\pi) \right| > \frac{m\, c_{0} }{k\, x_0} \frac{c_{0s}}{c_{0}} =
\gamma_0 \frac{c_{0s}}{c_{0}}
\]
where $c_{0s}$ is static friction coefficient. The first maxima/minima ($v=0$) in amplitude satisfying
\[ 
\left|x(n_{max}\pi) \right| > \gamma_0 \frac{c_{0s}}{c_{0}} \qquad
\left|x((n_{max}+1)\pi) \right| < \gamma_0 \frac{c_{0s}}{c_{0}} 
\]
gives the amplitude at which the motion is frozen.
So, in terms of energy using Eq. \eqref{eq:envelope0_E_Coulomb}, the condition for motion to continue after stopping 
($\mathcal{E}_{m} = \mathcal{E}_{p}$ at the amplitude maxima/minima) reads:
\[ 
\mathcal{E}_{m} (n \pi) = \left(1 - 2 n \gamma_0 \right)^2 = x^2(n\pi)
> \left( \gamma_0 \frac{c_{0s}}{c_{0}} \right)^2  
\]
limiting possible $n$ values from above.
This result enables us to determine maximum $n_{max}$, and, therefore, last half time oscillation
interval before stopping at $\tau=(n_{max}+1)\pi$. It is straightforward to show that motion
continue after stopping if
\[ n < \frac{1}{2\gamma_0} - \frac{c_{0s}}{2 c_{0}} \]
so we have arrived at the final expression:
\[ 
n_{max} = \left[\frac{1}{2\gamma_0} - \frac{c_{0s}}{2 c_{0}}\right] 
\]
must hold ($[]$ is the so-called floor operator, giving the greatest integer less than or equal to the argument). Now we can write down estimate of the residual potential energy for the DHO in the Coulomb case consistent with our approximation Eq. \eqref{eq:C_Em_app_g}:
\[
\mathcal{E}_{m} (\tau) = \left[ 1 - 2 (n_{max}+1) \gamma_0 \right]^2 \quad \tau \ge (n_{max}+1)\pi
\]
so we can write down an approximation for the energy:
\begin{flalign}
	n_{max} &= \left[\frac{1}{2\gamma_0} - \frac{c_{0s}}{2 c_{0}}\right] \nonumber \\
	\mathcal{E}_{m} (\tau) & = 
	\begin{cases}
		\left[1 - \gamma_0 \left(2n+1 - (-)^n \cos(\tau) \right)\right]^2 
		\qquad n = [\frac{\tau}{\pi}], & \text{if } \tau < (n_{max}+1)\pi \\
		\left[ 1 - 2 (n_{max}+1) \gamma_0 \right]^2, & \text{if }\tau \ge (n_{max}+1)\pi
	\end{cases} 
	\label{eq:coulomb_e_full}
\end{flalign}
In all the figures presented in this paper, for simplicity, we assume $c_{0s} = c_{0}$. 
Eq. \eqref{eq:coulomb_e_full} implies that observing the phenomenon where static friction exceeds kinetic friction necessitates a $c_{0s}/c_{0}$ ratio of at least 2 (as has been shown in \cite{Marchewka2004Coulomb} with somewhat more complicated derivation), given that this threshold is critical for the reduction of $n_{max}$.

If the Coulomb friction is present only, time between stopping instants is the same as for the non-damped case. Generally, in a weak-damping scenario, the approximation for oscillation period $T \simeq 2\pi / \omega_0$ is usually justified for any of three damping modes described here.
Fig. \ref{fig:Ek_r_E0_Coulomb} shows a comparison of Coulomb-damped oscillation periods for the kinetic and potential energy maxima/minima instants with those of the undamped HO, where time shifts are visible.

\subsection{Stokes damping \label{ssec:D_Stokes}}
Let us proceed along the same lines as in the previous case for Coulomb damping, now for Stokes damping. As in the Coulomb case, Stokes is exactly solvable, as we demonstrate in a novel way in the App. \ref{app:DHO_lin_sol}. Starting with \eqref{eq:e_tot_rate}
and \eqref{eq:EkEm_ratio} in this case we have:
\begin{equation} \label{eq:S_Ek_r_Em}
	\frac{d \mathcal{E}_{m}}{d\tau} =  
	- 2\gamma_1\, \mathcal{E}_k \simeq  - 2\gamma_1\, \mathcal{E}_m \sin^2(\tau)
\end{equation}
where we proceed similarly to the Coulomb damping case \eqref{eq:Ep_r_Em} solving first order equation by variable separation
\[
\frac{d \mathcal{E}_{m}}{\mathcal{E}_{m}} = -2\gamma_1 \sin^2(\tau) \,d\tau
\]
Upon integrating, we arrive at:
\[
\ln \mathcal{E}_{m} - \ln (\mathcal{E}_0) = -\gamma_1 \left[\tau - \frac{1}{2} \sin(2\tau) \right] 
\]
and finally
\begin{equation} \label{eq:S_en_app}
	\mathcal{E}_{m}(\tau) = \mathcal{E}_0 e^{-\gamma_1\tau } e^{\frac{\gamma_1}{2} \sin(2\tau) }
	= e^{-\gamma_1\tau } e^{\frac{\gamma_1}{2} \sin(2\tau) }
	\simeq e^{-\gamma_1\tau } \left[1 + \frac{\gamma_1}{2} \sin(2\tau)\right] 
\end{equation}
where initial condition $1 = \mathcal{E}_{m}(0)$ is assumed along with low damping limit $\gamma_1 \ll 1$.
In present context, weak damping means
\[ \frac{\gamma_1}{2} |\sin(2\tau)| \ll 1 \]
giving us approximation to the exact result with appropriate initial condition, see expression \eqref{eq:Em_Stokes}.
Given that Stokes' damping is a foundational concept in nearly all undergraduate reference materials \cite{Young2020university,HallidayResnick10} and is consequently subjected to thorough scrutiny and pedagogical treatment across academic levels, we avoid further examination of the topic in this study.

\begin{figure}[t]
	\includegraphics[height=7cm, angle=-0]{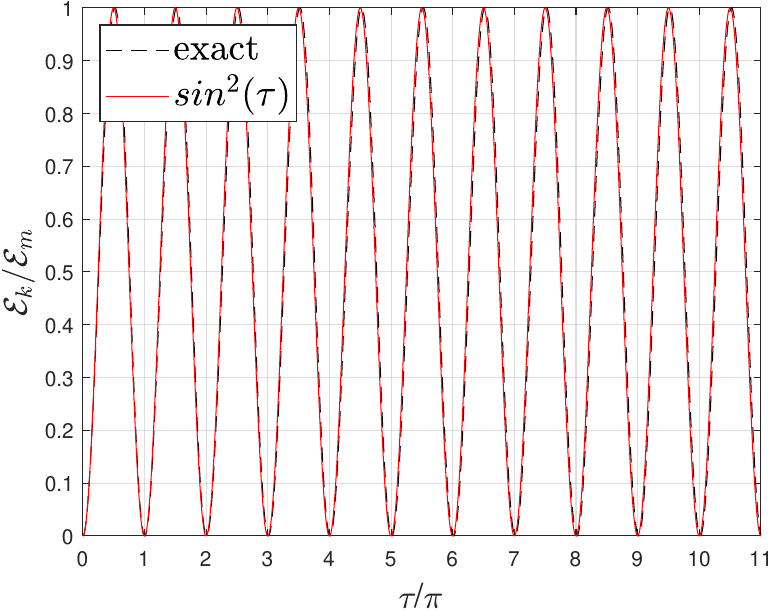}
	\includegraphics[height=7cm, angle=-0]{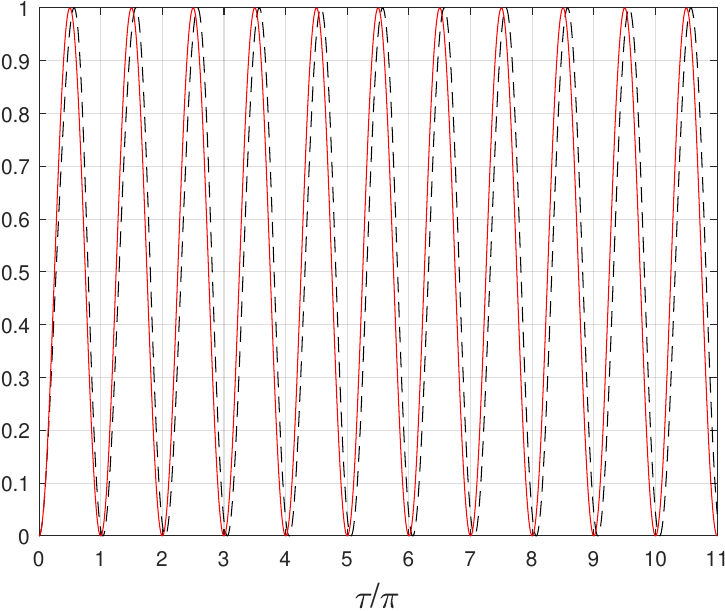}
	\caption{Newton damping, kinetic to total mechanical energy ratio, comparison of the exact (numerical) result (dashed black line) and approximation (solid red line) representing Eq. \eqref{eq:EkEm_ratio} for damping strengths $\gamma_2 = 0.20$ (left pane) $\gamma_2 = 0.60$ (right pane). \label{fig:Ek_r_E0_Newton}}
\end{figure}
\subsection{Newton damping \label{ssec:D_Newton}}

\begin{figure}[h!]
	\includegraphics[width=0.45\textwidth, angle=-0]{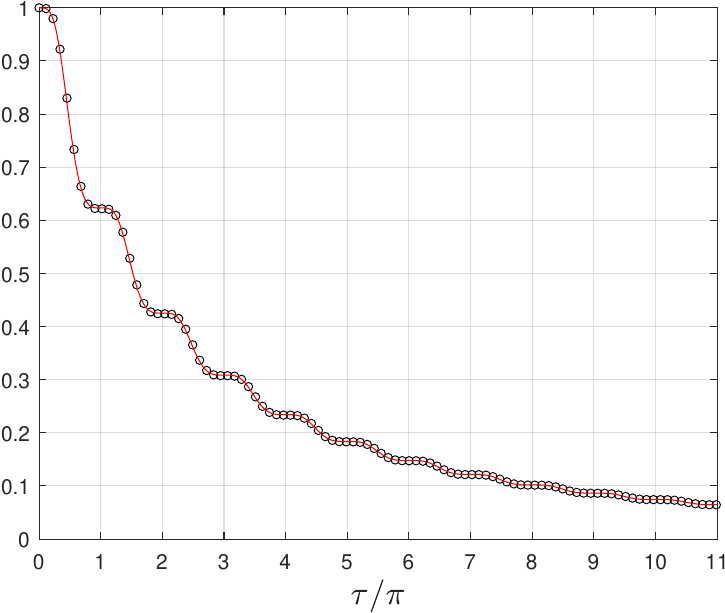}
	\includegraphics[width=0.45\textwidth, angle=-0]{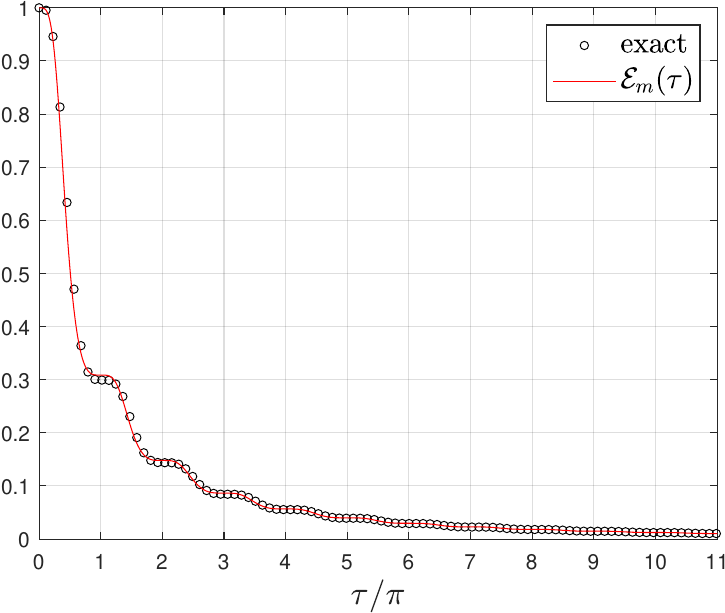}
	\caption{Newton damping, mechanical energy decay exact (numerical, circles) and approximation (solid red line) Eq. \eqref{eq:N_Em_app_g} for damping strengths $\gamma_2 = 0.20$ (left pane) $\gamma_2 = 0.60$ (right pane). \label{fig:Etot_Newton}}
\end{figure}

Proceeding along the same lines, as in the previous cases of Coulomb and Stokes damping, for Newton damping, we start using general expressions with \eqref{eq:e_tot_rate} and \eqref{eq:EkEm_ratio}:
\begin{equation} \label{eq:N_Ep_r_Em}
\frac{d \mathcal{E}_{m}}{d\tau} =  
- 2\gamma_2\, \left(\mathcal{E}_k \right)^\frac{3}{2} \simeq  - 2\gamma_2\, \left(\mathcal{E}_m \right)^\frac{3}{2} | \sin(\tau) |^3
\end{equation}
analogously to Eq. \eqref{eq:Ep_r_Em}, we arrive at the expression ready for solving by separation of variables
\[
	-\frac{d \mathcal{E}_{m}}{2 \mathcal{E}_{m}^{\frac{3}{2}}} = \gamma_2 | \sin(\tau) |^3 \,d\tau
\]
Following the same solution strategy (solving on intervals where $\sin(\tau)$ is either positive or negative and ensuring continuity of the solution) we derive an approximate expression for mechanical energy in the case of Newton damping as well:
\begin{equation} \label{eq:N_Em_app_g}
	\mathcal{E}_{m} (\tau) = \left[1 + \frac{2}{3} \gamma_2 
	\left(2n+1 + \frac{(-)^n}{2} \left(cos^3(\tau) - 3\cos(\tau)\right) \right)\right]^{-2} \qquad 
	\tau \in \pi [n,n+1], n=0,1,2,3,\ldots
\end{equation}
Analogously to the Coulomb damping Fig. \ref{fig:Ek_r_E0_Newton} shows approximation, Eq. \eqref{eq:EkEm_ratio}, for Newton damping, while Fig. \ref{fig:Etot_Newton} illustrate accuracy of the Eq. \eqref{eq:N_Em_app_g} for two values of the damping strengths.

Similarly to the case with Coulomb damping, the approximate total energy Eq. \eqref{eq:N_Em_app_g} consists entirely of the potential energy\footnote{Assertion is an approximation; exact $\mathcal{E}_{k} (\tau_n) = 0$ instants arrive slightly later in a weak damping limit as clear from Fig. \ref{fig:Ek_r_E0_Newton}.} at instants $\tau_n=n\pi$, for $n\in\lbrace0,1,2,\dots\rbrace$. Using $\tau=\tau_n$ and $\cos(\tau_n)=(-1)^n$ in \eqref{eq:N_Em_app_g}, we easily get
\begin{equation} \label{envelopa00}
	\mathcal{E}_{m} (\tau_n) = \left(1 + \frac{4n}{3}\,\gamma_2\right)^{-2}=\left(1 + \frac{4\gamma_2}{3\pi}\,\tau_n\right)^{-2}\,,
\end{equation}
where (again) we formally wrote $n=\tau_n/\pi$ after the second equality to explicitly show how the maxima of the potential energy decay over time. Thus, we can easily conclude that the envelope of the potential energy as a function of time is approximately given by 
\begin{equation*} 
	\mathcal{\tilde{E}}_{p} (\tau) = \left( 1 + \frac{4}{3\pi} \gamma_2 \,\tau \right)^{-2}\,.
\end{equation*}
\begin{figure}[b]
	\includegraphics[width=0.45\textwidth, angle=-0]{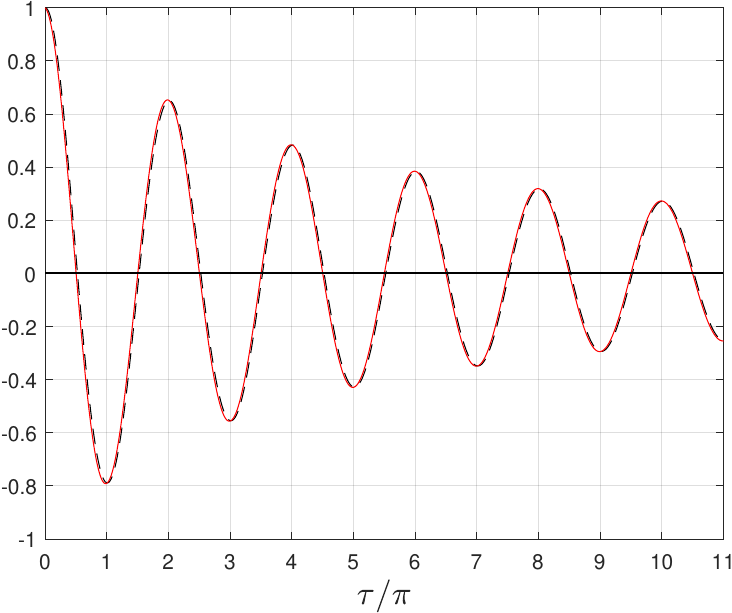}
	\includegraphics[width=0.45\textwidth, angle=-0]{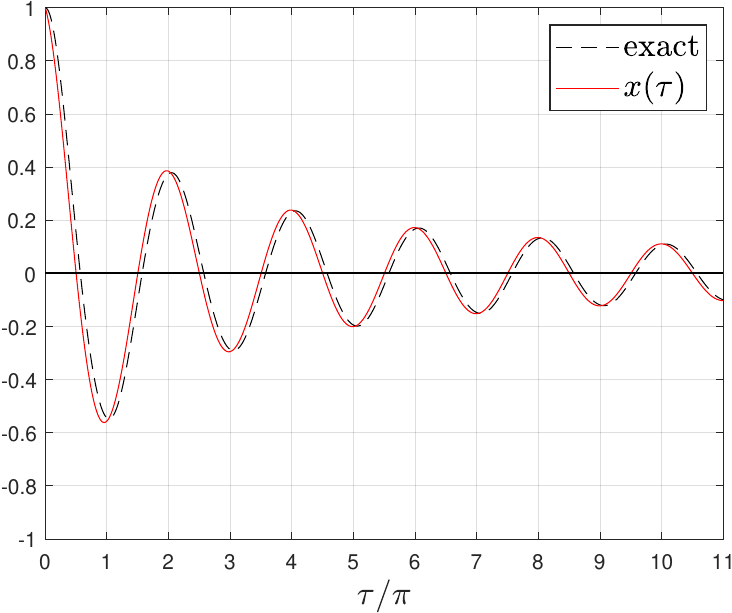}
	\caption{Newton damping, exact (numerical, dashed black line) and approximate displacement (solid red line) Eq. \eqref{eq:x_app_Coulomb} for damping strengths $\gamma_2 = 0.20$ (left pane) $\gamma_2 = 0.60$ (right pane). \label{fig:x_Newton}}
\end{figure}
The same argumentation as in the case of Coulomb damping leads to
\begin{equation}
	x(\tau)=\frac{\cos(\tau)}{1 + \frac{4}{3\pi} \gamma_2 \,\tau}
	\label{eq:x_app_Coulomb}
\end{equation}
as an approximation of displacement as a function of time shown on Fig. \ref{fig:x_Newton} for two damping strengths, and to
\begin{equation} \label{eq:approxEm_Newton}
	\mathcal{\tilde{E}}_{m} (\tau) = \left(1 + \frac{4}{3\pi} \gamma_2 \,\tau\right)^{-2}\,
\end{equation}
as a closed-form approximation of the total mechanical energy, we note that expression \eqref{eq:approxEm_Newton} was already derived in \cite{WangEJP2002}, but we derived it here without the averaging procedure.

Fig. \ref{fig:Etot_Newton} shows the comparison of the energy approximation obtained here Eq. \eqref{eq:N_Em_app_g} and the exact (numerical) solution, while Fig. \ref{fig:x_Newton} shows the comparison of the amplitude calculated using the approximation and the numerically calculated one.

Approximation is motivated by weak damping limit considerations, but a quantitative examination shows solid agreement even out of the weak damping limit (exemplified here with damping strength $\gamma_2 = 0.60$). For a more detailed precision analysis see \cite{LelasPezer2025Part1}. As a rule of thumb, approximation is valid if the system goes through at least several oscillations. For Newton damping, the approximation is valid over a wide range of damping coefficients. The reason is simple, damping quadratic in velocity, even if strong in the beginning, effectively ends up in a weak damping regime after one or two oscillation cycles.

\section{Benefits of our results in undergraduate physics teaching \label{sec:benefits}}

Our derivation of the solution for a harmonic oscillator with linear damping, presented in Appendix \ref{app:DHO_lin_sol}, is designed to be comprehensible even to students unfamiliar with conventional methods for solving second-order differential equations. Furthermore, envision a cohort of students proficient in the analytical solutions and dynamics of a simple harmonic oscillator, as well as the fundamentals of differential calculus, yet lacking formal prior knowledge of a damped harmonic oscillator, except for having seen an experiment demonstrating free oscillations in a weakly damped regime. In that case, our approach leads them to an exact solution for the underdamped regime without any insight into the existence of critical and overdamped regimes (since it is not necessary to solve the characteristic polynomial, which in the standard approach indicates the existence of all three regimes from the start). Having obtained the underdamped solution, students can consider its behavior with increasing damping strength and theoretically predict the existence of a critical ($c_1/2=\omega_0$) and overdamped ($c_1/2>\omega_0$) regimes and obtain the corresponding solutions easily, see, e.g., Section VI in \cite{LelasPezerEJP2024Mod}. Finally, students can experimentally verify their theoretical predictions. 

The scheme presented precisely exposes one way in which physics progresses. For example, we found a good theoretical description of experiments performed with a range of values of some parameter that was available at the time (in our example, e.g. $0<c_1/2\leq0.2\omega_0$), and that theoretical description gives us a prediction of qualitatively different behavior for values of parameter that are far beyond the reach of experimental possibilities at that time (e.g. for $c_1/2\gtrsim 0.9\omega_0$). Should these theoretical predictions prove compelling, experiments will be designed to achieve higher values of the relevant parameter (e.g., $c_1/2\gtrsim 0.9\omega_0$), enabling empirical verification or falsification of the underlying theoretical description. Therefore, this simple system can serve to illustrate to students the interplay between theory and experiments in physics, leading to a better understanding and/or better modeling of the phenomena.

Our approximate description of the energy decay also brings an important lesson, i.e., with elementary insights into the physics of the problem, approximations can be made that significantly simplify even analytically unsolvable issues (as a harmonic oscillator with quadratic damping) and at the same time provide a good quantitative description of the system's behavior. Furthermore, based on our expressions for energy decay, i.e., relations \eqref{eq:C_Em_app_g}, \eqref{eq:S_en_app}, and \eqref{eq:N_Em_app_g}, engaging laboratory exercises can be designed. Namely, using tracker software \cite{BrownCox2009}, one can experimentally obtain position vs. time and velocity vs. time data, e.g., for a block-spring system with sliding friction \cite{Kamela2007Coulomb}, a physical pendulum with quadratic damping \cite{WangEJP2002}, or some other system described by the damped harmonic oscillator model. From these experimental data, it is possible to plot curves corresponding to the experimentally determined energies and compare them with curves obtained using our approximate energy-decay expressions.

\section{Conclusion \label{sec:conclusion}}

By comparing our analytical treatment with exact/numerical solutions, we have obtained an excellent and straightforward description of the decay modulation of the damped HO in the Coulomb and Newton drag regimes. In addition to the Stokes drag case, typically presented in undergraduate physics literature, we provided a simple derivation of an exact solution of the underlying differential equation. In summary, we have shown, from the energy point of view, key characteristics of the three forms of dissipation in the harmonic oscillator case:
\begin{enumerate} 
	\item Coulomb (Constant) Damping, force law: $F_d = -m c_0\, sgn(v)$, independent of speed.
	\begin{itemize}
		\item Effect: Kinetic/potential energy local maxima decrease quadratically with time, see Eq. \eqref{eq:approxEm_Coulomb}; motion ceases after a finite time.
		\item Key feature: Nonlinear, discontinuous force; amplitude drops by a fixed amount each half-cycle. The half oscillation period time (the time between subsequent stops) is the same as in HO without damping.
	\end{itemize}
	\item Stokes (Linear) Damping, force law: $F_d = -m c_1\, v$.
	\begin{itemize}
		\item Effect: Kinetic/potential energy local maxima decrease exponentially, see Eq. \eqref{eq:Em_Stokes}.
		\item Key feature: Frequency shift is small ($\omega_d \simeq \omega_0$) for weak damping. Oscillation period is constant; damping introduces a phase lag but preserves sinusoidal character.
	\end{itemize}
	\item Newton (Quadratic) Damping, force law: $F_d = -m c_2\, sgn(v)\, v^2$.
	\begin{itemize}
		\item Effect: Kinetic/potential energy local maxima decrease as inverse square in time, see Eq. \eqref{eq:approxEm_Newton}.
		\item Key feature: Dissipation is strongest at high speeds; a modest increase in the first few semicycles' length relative to the undamped HO is observed in a weak-damping regime at large amplitudes, stemming from asymmetric slowing near zero crossings.
	\end{itemize}
\end{enumerate}
The only approximation used in formulating and solving the appropriate equations is the kinetic-to-total mechanical energy ratio (a positive, oscillatory function with values in the interval [0,1]), taken from undamped HO. It has been demonstrated that the unified approach presented here substantially simplifies the underlying formalism for all three damping modes while remaining very precise, maintaining high precision even with considerable damping forces, and preserving almost all the physical effects of interest across diverse educational contexts. In the case of Newton damping, where the analytical framework becomes overly complex \cite{Smith2012Newton} and hinders introduction in undergraduate education, the current approach offers novel opportunities.

Including Coulomb (constant) and Newton drag in an introductory physics course provides clear pedagogical benefits by broadening students’ understanding of damping beyond idealized linear models. It exposes students to a physically realistic, nonlinear force law and demonstrates that exponential decay is not universal for dissipative systems. Coulomb drag naturally emphasizes work–energy reasoning over differential equation solving, showing that a constant friction force removes a fixed amount of energy per unit distance traveled, leading to linear amplitude decay and quadratic energy decay. Finally, it explains the finite stopping time and abrupt cessation of motion observed in everyday mechanical systems, strengthening physical intuition and reinforcing the importance of choosing models based on dominant physical effects rather than mathematical convenience. Newton drag also supports energy-based reasoning in nonlinear dynamics, featuring swift dissipation in the first one or two cycles and a gradual reduction in energy afterward.

In-class experimental demonstrations for all three types of drag considered here are relatively easy to prepare. Experiments further reinforce these pedagogical advantages by allowing students to observe the distinctive signatures of Coulomb and Newton drag directly. Simple setups such as a mass–spring system on a rough surface or a pendulum with a friction pad exhibit linear amplitude decay and a finite stopping time, in stark contrast to the smooth exponential decay observed with Stokes (air or viscous) damping. By measuring peak displacements over successive cycles, students can connect theoretical energy arguments to empirical data without demanding mathematical analysis. These demonstrations make the abstract concept of diverse drag versions tangible, highlight the predictive power of energy-based reasoning, and deepen students’ intuition about how real mechanical systems dissipate energy.

\begin{acknowledgments}
	R.P. and K.L. acknowledge support from the project “Implementation of cutting-edge research and its application as part of the Scientific Center of Excellence for Quantum and Complex Systems, and Representations of Lie Algebras”, Grant No. PK.1.1.10.0004, co-financed by the European Union through the European Regional Development FundCompetitiveness and Cohesion Programme 2021-2027.
\end{acknowledgments}

\appendix
\section{Stokes damping, exact treatment employing an ansatz \label{app:DHO_lin_sol}}

Second Newton's law for the HO with linear only damping in velocity Eq. \eqref{eq:Newton2nd4x} reads:
$$\ddot{x}_d +  {c_{1}}\dot{x}_d + \omega_0^2 \, x_d = 0$$
Here $c_1 = b/m$, connecting present to the usual notation as in \cite{HallidayResnick10}.
Now, we'll assume that the solution, motivated by experimental observations, is of the form:
$$x_d(t) = f(t) \, cos(\omega_d t +\phi)$$
where $\omega_d$ is the unknown constant (we allow for an angular frequency different from $\omega_0$) and amplitude f(t), the unknown function, to be determined. In systems where an elastic restoring force coexists with drag, the observed oscillatory motion appears as a combination of two parts: HO oscillations modulated by a positive, monotonically decreasing function. An oscillator system with negligible drag is challenging to realize in an undergraduate LAB setting (e.g., the torsion oscillator); understanding damping is therefore fundamental from a physics perspective. Since in standard undergraduate physics textbooks \cite{HallidayResnick10, Young2020university} the damped harmonic oscillator is presented by stating the exact solution for the linear damping case, we consider this a missed educational opportunity, as we show here. The ansatz approach is no more technically challenging than checking the validity of the exact form, but, as a bonus, it introduces practical techniques and reinforces physical intuition.

For $f(t)$, we expect a monotonically decaying function according to experimental observations.
We will simplify algebraic manipulations by setting $\phi=0$, since it is influenced only by the initial conditions. At the end, we will add a phase to the general solution.
Substituting this ansatz into the equation, we get:
\[
0 = \frac{d^2}{dt^2}(f(t) \, cos(\omega_d t)) + {c_{1}}\frac{d}{dt}(f(t) \, cos(\omega_d t)) 
+ \omega_0^2 \, f(t) \, cos(\omega_d t)
\]
Using the product rule for differentiation, we can expand the derivatives:
\begin{equation*}
	0 = \ddot{f}(t) cos(\omega_d t) - 2\omega_d \dot{f}(t) sin(\omega_d t) - f(t) \, \omega_d^2 cos(\omega_d t) 
	+ {c_{1}} \dot{f}(t) cos(\omega_d t) - {c_{1}} f(t) \, \omega_d  sin(\omega_d t) 
	+ \omega_0^2 \, f(t) \, cos(\omega_d t)
\end{equation*}
Simplifying and rearranging, we get:
\begin{equation} \label{eq:diff4f_cond}
	0 = \cos (\omega_d t ) \left(f''(t)+ {c_{1}}  f'(t)
	+\left(\omega_0^2-\omega_d^2\right) f(t)\right) \\
	-2 \omega_d  \sin (\omega_d t ) \left(f'(t)+{\frac{c_{1}}{2}} f(t)\right)
\end{equation}
We can again rely on the linear independence of the functions $sin(\omega_d t)$ and $cos(\omega_d t)$, what implies that the coefficients for both $sin(\omega_d t)$ and $cos(\omega_d t)$ must be equal to zero since \eqref{eq:diff4f_cond} holds for any $t$. Therefore, when we apply the ansatz $x(t) = f(t) \, cos(\omega_d t)$, the damped harmonic oscillator differential equation transforms into:
\begin{subequations} \label{eq:diff4f}
	\begin{align}
		\omega_d \dot{f}(t) + \omega_d   {\frac{c_{1}}{2}} f(t) = 0,\, \omega_d \ne 0 \, \Longrightarrow  \, \dot{f}(t) + {\frac{c_{1}}{2}} f(t) = 0
		\label{eq:diff4f_1}\\
		\ddot{f}(t) + c_{1} \dot{f}(t) + (\omega_0^2 - \omega_d^2) f(t) = 0  \label{eq:diff4f_2}
	\end{align}
\end{subequations}
Eq. \eqref{eq:diff4f_1} is a first-order linear ordinary differential equation (ODE) for f(t), while Eq. \eqref{eq:diff4f_2} is a second-order linear ODE for f(t).
If we assume $\omega_d \ne 0$, solving \eqref{eq:diff4f_1} is straightforward since the equation is a unique property of the exponential function, we have:
\begin{equation} \label{eq:f_sol}
	f(t) = A \, e^{- {\frac{c_{1}}{2}} t}
\end{equation} 
where $A$ is an arbitrary constant. By substituting this solution into Eq. \eqref{eq:diff4f_2}, we can assess the consistency of the differential equations and obtain the following condition:
\[
0 =
\ddot{f}(t) + c_{1}  \dot{f}(t) + (\omega_0^2 - \omega_d^2) f(t) 
= \left[\left({\frac{c_{1}}{2}}\right)^2 - c_{1}^2 + \omega_0^2 - \omega_d^2 \right] A\, e^{- {\frac{c_{1}}{2}} t}
\]
Again, this holds for any $t$, so we end up in an algebraic equation:
$$ - \frac{c_{1}^2}{4} + \omega_0^2 - \omega_d^2 = 0 $$
which is simply a constraint on the model parameters:
\begin{equation} \label{eq:dho_v_omd}
	\omega_d^2 = \omega_0^2 -  \frac{c_{1}^2}{4}
\end{equation}
relating unknown frequency $\omega_d$, HO natural frequency $\omega_0$, and drag coefficient $c_{1}$. We have successfully finalized the solution to the dynamic Eq. \eqref{eq:Newton2nd4x}, having sidestepped the process of solving a second-order differential equation, giving us the final solution in general form:
\begin{equation} \label{eq:dho_sol}
	x_d(t) = A\, e^{-{\frac{c_{1}}{2}} t} \cos(\omega_d t+\phi)
\end{equation}

In case of initial conditions $x_d(0)=x_0>0$ and $v(0)=0$, \eqref{eq:dho_sol} can be rewritten as \cite{LelasPezerEJP2024Mod, Crawford1968BerkeleyWaves}
\begin{equation}
	x_d(t)=x_0 e^{-\frac{c_1}{2}t}\left(\cos(\omega_d t)+\frac{c_1}{2\omega_d}\sin(\omega_d t)\right)\,.
	\label{eq:XDHOsolution}
\end{equation}
The velocity corresponding to \eqref{eq:XDHOsolution} is
\begin{equation}
	v(t)=\frac{dx(t)}{dt}=-\frac{\omega_0^2x_0}{\omega_d}e^{-\frac{c_1}{2}t}\sin(\omega_d t)\,.
	\label{eq:VDHOsolution}
\end{equation}
Using \eqref{eq:XDHOsolution} and \eqref{eq:VDHOsolution} we obtain the total mechanical energy  
\begin{equation}
	E_m(t)=E_0e^{-c_1t}\left(1+\frac{c_1}{2\omega_d}\sin(2\omega_d t)+2\left(\frac{c_1}{2\omega_d}\right)^2\sin^2(\omega_d t)\right)\,,
	\label{eq:Em_Stokes}
\end{equation}
where $E_0=m\omega_0^2x_0^2/2$. The corresponding kinetic energy is
\begin{equation*}
	E_k(t)=E_0e^{-c_1t}\frac{\omega_0^2}{\omega_d^2}\sin^2(\omega_d t)\,.
\end{equation*}
In the case $c_1\ll\omega_0$, we have $\omega_d\approx\omega_0$, and we can easily see that 
\begin{equation}
	\frac{E_k(t)}{E_m(t)}\approx\sin^2(\omega_0 t)\,
	\label{eq:EkEm_ratio_ApproxApp}
\end{equation}
is a valid approximation if $c_1/(2\omega_0)\ll1$ holds.

\bibliography{main}

\end{document}